\documentclass[prb,showpacs,preprintnumbers,amsmath,amssymb]{revtex4}

\usepackage{graphicx}% Include figure files
\usepackage{dcolumn}% Align table columns on decimal point
\usepackage{bm}% bold math

\begin{document}

%\preprint{APS/123-QED}
%\draft
%\twocolumn[\hsize\textwidth\columnwidth\hsize\csname@twocolumnfalse\endcsname

\title{Taming the Gerrymander---Statistical Physics Approach to Political Districting Problem}
\author{Chung-I Chou}
\affiliation{Department of Physics, Chinese Cultural University, Taipei, Taiwan 111, R.O.C.} 
%\author{Sai-Ping Li{\cite{li}}} 
\author{Sai-Ping Li}
\affiliation{Institute of Physics, Academia Sinica, Taipei, Taiwan 115, R.O.C.}
\date{\today}
%\maketitle

\begin{abstract}
The Political Districting Problem is mapped to a $q$-state Potts model in which the constraints 
can be written as interactions between sites or external fields acting on the system.  Districting
into $q$ voter districts is equivalent to finding the ground state of this $q$-state Potts model.  
We illustrate this by districting Taipei city in its 2008 Legislature Election.
Statistical properties of the model are also studied.  

\end{abstract}

\pacs{89.65.Cd; 89.65.Ef; 89.65Gh; 89.90.+n}
%\vspace{2pc}]

\maketitle

\section{Introduction}

In 1812, Massachusetts governor Elbridge Gerry, got help from his political party
by crafting a district and won his own election.  At the time, someone produced an illustration
of the districting and emphasized its similarities with a salamander.  The term 
{\it {Gerrymander}} was then coined from putting together Gerry and mander.  Nowadays, 
{\it{Gerrymandering}} refers to the practice of drawing district lines to maximize the 
advantage of a political party.  For example, a bipartisan gerrymandering is the one in which
the districting is to protect incumbents, and a racial or ethnic gerrymandering is to dilute or
preserve the strength of minorities.  

Political Districting has since become an issue that is always political, controversial and
sometimes even ugly.  In the US, for example, the results of a population census in every
ten years may require a voter redistricting in order to redistribute the House seats among the
states.  Politicians have always fought over district boundaries, while the courts might
consider the problem just too political even to enter.  Several constitutional amendments 
were actually passsed in the nineteenth century to prevent {\it{Gerrymandering}}.  In
nearly half of the states that underwent voter redistricting in the 1990s, federal or state
courts played an essential role in the redistricting debate and judges actually issued new 
lines in ten states.  In this process of redistricting, the courts indeed never used quantitative 
methods to justify the plans.  One would wonder if there are more objective methods to
perform the redistricting process.

There are actually mathematical and numerical approaches exist in the literature.  Such
methods can in principle eliminate {\it{Gerrymandering}} by providing well defined
steps and constraints.  Local search methods include those used in Kaiser [1]
 and in Nagel [2].  An implicit enumeration technique was also developed by 
Garfinkel and Nemhauser [3].  George et.al. [4] studied the problem of 
determining New Zealand's electoral districts, using a location-allocation based
iterative method in conjunction with a geographic information system (GIS).  

From a mathematical point of view, the Political Districting Problem belongs to what is known 
as the Districting (or zone design) Problem in which $n$ units are grouped into $k$ 
zones such that some cost function is optimized, subject to constraints on the topology 
of the zones, etc and has been shown to be NP-Complete [5].  Thus, it is best to be
treated by some optimization methods.  
The Districting Problem is a geographical problem which is present in
a number of geographical tasks such as school districting, design of sales territories, etc.  
The constraints of the Districting Problem are very similar to that of the Clustering Problem 
in optimization.  Let the set of $n$ initial units be $X = {x_1, x_2, ..., x_n}$, where $x_i$
is the $i$-th unit and let the number of districts be $k$.  Let $Z_i$ be the set of all the units 
that belong to district $Z$.  Then

$$
Z_i \neq \emptyset \,\,\, , i = 1, ..., k \,\,\, , 
$$
$$
Z_i  \cap Z_j = \emptyset \,\,\, , i \neq j \,\,\, ,
$$
$$
\cup^k_{i=1}  Z_i = X \,\,\, .
\eqno(1.1)
$$

There is an additional constraint in the Districting Problem, namely, the constraint
of contiguity which makes the problem somewhat more complicated.  It constrains the set 
of possible solutions to the problem that assures contiguity between the units within the 
designed district.  Contiguity here means that every unit in a district is connected to every 
other unit through units that are also in the district.  An important optimization criterion 
in the Political Districting Problem is to avoid Gerrymandering.  It is generally accepted 
that there are three essential characteristics that the districts should have [6]: 
population equality, contiguity and geographical compactness.  The task here is 
therefore to devise a method that is able to produce solutions which satisfy these 
characteristics.  

In this paper, we map the Districting problem onto a $q$-state Potts model.  This would
allow us to study the problem by using statistical physics methods.  Most of the constraints
that we mentioned above could be represented as the interaction terms among various sites
of the $q$-state Potts model or the addition of an external field to the system.  By doing so, 
we can also understand the corresponding physical nature of such a social problem.  

Using a physics model to study a social science problem is not new.  
There already appear many papers and books written on various 
subjects in social science.  People have employed
concepts such as scalings, etc, to study the social behavior in financial markets [7-9].  
Statistical models have also been applied to NP-complete 
problems in combinatorial optimization [10].  We here demonstrate how a social 
economics problem can be transformed into a physics model and carry out an optimization 
study to look for the optimal solution of the problem.  This paper is organized as follows.  
Section II is a description of the model for the redistricting problem.  Section III contains 
the results of our numerical simulation and Section IV is the summary and discussion.

\section{The Model}

The $q$-state Potts model was first proposed as an appropriate generalization of the
Ising model, to consider a system of spins confined in a plane with each spin pointing 
to one of the $q$ equally spaced directions specified by the angles

$$
\Theta_n = \frac{2\pi n}{q} \,\,\, , n = 0, 1, ..., q-1 \,\,\, .
\eqno(2.1)
$$

\noindent
In the most general form, the nearest-neighbor interaction would only depend on the 
relative angle between the two vectors.  The Hamiltonian will then take the form

$$
H = - \sum_{ij} J(\Theta_{ij}) \,\,\, , 
\eqno(2.2)
$$

\noindent
where the function $J(\Theta)$ is $2\pi$ periodic and 
$\Theta_{ij} = \Theta_{n_i} - \Theta_{n_j}$ is the angle between the two spins of
neighboring sites $i$ and $j$.  

In his seminal work, Potts [11] chose

$$
J(\Theta) = \epsilon_1 \cos\Theta \,\,\, ,
\eqno(2.3)
$$

\noindent
and was able to determine the critical point of this (now known as planar Potts) model on 
the square lattice for $q = 2, 3, 4$.  As a remark, he also gave the critical point for all $q$
of the following (now known as standard Potts) model

$$
J(\Theta_{ij}) = \epsilon_2 \delta_{S_i, S_j} \,\,\, ,
\eqno(2.4)
$$

\noindent
where $\delta_{ij}$ is the Kronecker delta and is equal to 1 when $i=j$ and 0 otherwise.  
It is also the model with interaction energy of 
the form in Eq. (2.4) that has attracted the most attention to date.  In our study here, 
we shall use this standard Potts model as the starting point.  

To begin with, we assume each site, or $q$-state spin to represent a unit in the Districting 
problem and with a total of $N$ units.  $q$ is the total number of districts in the plan.
For a spin to be in one of the $q$ states means that the unit belongs to that particular district. 
Define $n_j$ to be the number of sites in a particular state $j$.  It is clear that 

$$
\sum_{j=1}^q n_j = N \,\,\, .
\eqno(2.5)
$$

\noindent
The objective here is to include the constraints such as population equality, contiguity 
and geographical compactness as interaction energy terms in the Hamiltonian so that the
ground state energy configuration corresponds to the optimal solution of the problem.  

Achieving equal voter population size districts is central to any Political Districting Problem.  
A measure of this can be obtained, e.g., by calculating the sum of the differences 
between the population of each district and the average population over all districts.  
To include this in our model, we can view it as an external random field acting on
site $i$ with a field strength $p_i$ representing the voter population of this site.  
Therefore, the total voter population $P_j$ of a district $j$ is the sum of the interaction of
the external random field and all the sites within this district and is given by

$$
P_j = \sum_{i=1}^{n_j} p_i \delta_{S_i, j} \,\,\, .
\eqno(2.6)
$$

\noindent 
and the total voter population of the plan is given by

$$
P_0 = \sum_{j=1}^q P_j \,\,\, .
\eqno(2.7)
$$

\noindent
The average voter population $\langle P \rangle$ for each district is therefore equal to 
$P_0/q$.  The difference between the voter population of district $j$ and the average 
voter population will contribute to the Hamiltonian of the system.  Let us define its 
total contribution to be $H_P$.  It is then given by

$$
H_P = \sum_{j=1}^q  \biggl| \frac{P_j}{\langle P \rangle} - 1  \biggr| \,\,\, .
\eqno(2.8)
$$

Population equality alone will sometimes lead to problems of contiguity and compactness
in districting, resulting in districts of unnatural shapes.  Hence compactness is usually 
an important factor in any political districting solution.  There are many ways to define
the compactness of a district but there is yet no universally acceptable definition of 
compactness.  Young [12] studied eight different measures of compactness
and showed that each measure fails to give satisfactory results on certain geographical
configurations.  In short, any good measure of compactness must apply both to the
districting as a whole and to each district individually.  It should also be conceptually
simple and should use easily collected and verifiable data.  Our strategy here is to take
compactness as the smallest total sum of all boundaries between different districts.  In
this way, we can view it as the interaction energy between domains, or the domain 
wall energy.  Thus, the contribution between sites $i$ and $j$ to the domain wall
energy is

$$
( 1 - \delta_{S_i, S_j} ) C_{ij} \,\,\, ,
\eqno(2.9)
$$

\noindent
where $C_{ij} = 1$, if $i$ and $j$ are neighboring sites and zero otherwise.  Define
the Hamiltonian energy from this interaction to be $H_D$, we will have

$$
H_D = \sum_{i,j} ( 1 - \delta_{S_i, S_j} ) C_{ij} \,\,\, .
\eqno(2.10)
$$

$H_P$ and $H_D$ are the two energy terms that we include in our Hamiltonian for
the study of the system, which we now define as

$$
H =  \lambda_P H_P + \lambda_D H_D \,\,\, ,
\eqno(2.11)
$$

\noindent
where $\lambda_P$ and $\lambda_D$ are constant coefficients.  Notice that the way
we define $H_D$ would in most cases quarantee the constraint of contiguity, depending 
on the ratio between $\lambda_P$ and $\lambda_D$.  Other constraints can also be
included as the interaction among various sites or external fields acting on the system
and will be discussed in Sec. IV.

\section{Numerical Simulation Results}

In the above section, we have shown how to map the Districting Problem to a 
$q$-state Potts Model and rewrite the constraints into interaction between different
sites and external fields acting on the system.   
In this section, we will use the problem of determining the districting for the Taiwan 
Legislature seats as an example, though the method can be equally applicable to any 
districting problem.  We will also use Monte Carlo method to perform our simulation.

Starting from the 2008 Legislature election, the political districts of Taiwan will be 
restructured, resulted in 73 voter districts where
each district will elect its own legislator for the Legislature.  
Taipei city, for example, will have 8 voter districts.  On the other hand, Taipei city 
has at this moment a total of 449 precincts (a precinct corresponds to a site in our model) 
and a voter population of about 2 million. 
Redistricting into 8 voter districts will imply that we need to regroup and put the 
precincts into different voter districts, which is a typical Political Districting Problem.  
In order to satisfy the population equality constraint, a voter population  
of about 250,000 voters in each voter district is preferred.  

Figure 1 shows the distribution of the number of precincts vs. precinct voter population 
in Taipei city.  The y-axis is the number of precincts while the x-axis is the voter 
population in each precinct, with each bin representing
500 voters.  It is interesting to see that this can be approximated by a Gaussian distribution 
with a mean of about 4,100 and a variance of about 1,450.  This can in turn 
be interpreted as a Gaussian-distributed random field term in the $q$-state Potts model.  
We should mentiont here that $q$-state Potts model with 
Gaussian-distributed random fields is of tremendous interest in physics.  It has been studied 
for a long time and will not be discussed here.  

%Figure 1
\begin{figure}
\includegraphics[width=10cm]{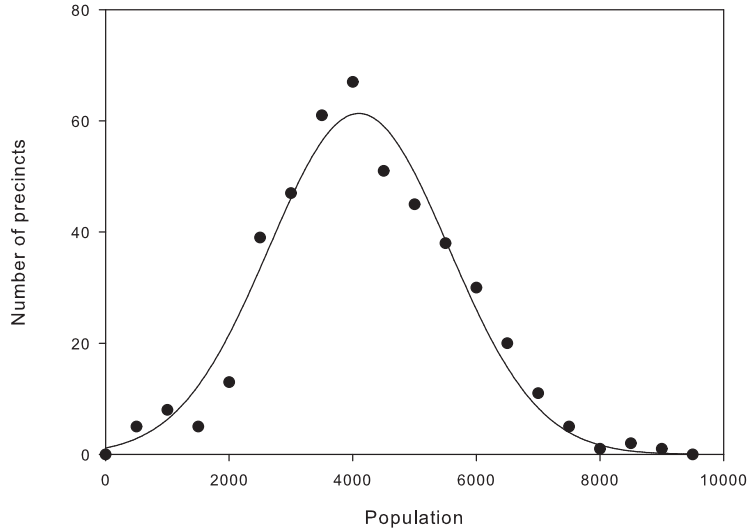}% Here is how to import EPS art
\caption{The distribution of the number of precincts vs. precinct 
voter population in Taipei city.}  
\end{figure}

%Figure 2
\begin{figure}
\includegraphics[width=12cm]{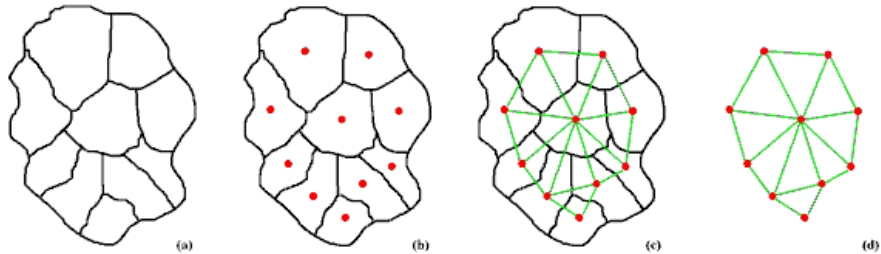}% Here is how to import EPS art
\caption{(a) A district of 10 precincts in our model; (b) each red dot represents one
precinct; (c) a network of precincts connected by green arcs; (d) the network 
extracted from (c).  }  
\end{figure}

%Figure 3
\begin{figure}
\includegraphics[width=10cm]{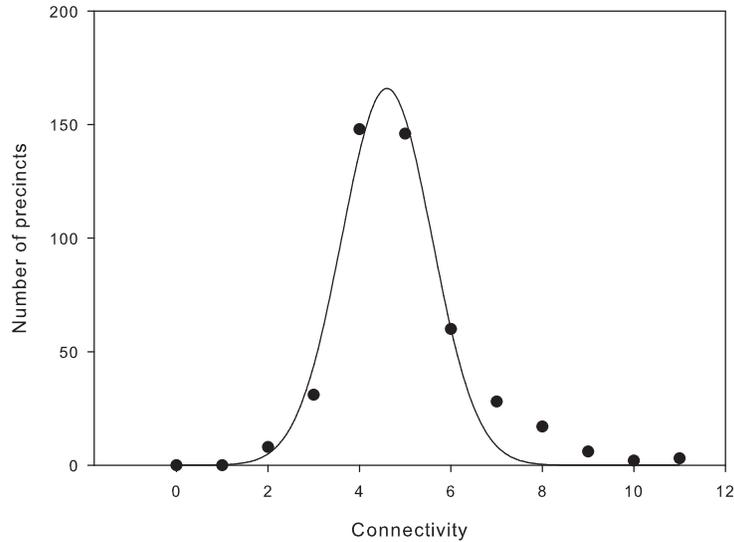}% Here is how to import EPS art
\caption{The distribution of the number of neighbors that each of the 
current 449 precincts in Taipei city has.}  
\end{figure}

In order to map this problem
onto the $q$-state Potts model, we need to define what is meant by the neighbors of a 
site (precinct).  Figure 2 is an illustration of our definition.  It represents a district with 10
precincts, as shown in Figure 2(a).  Each of the precincts is represented by a red dot, 
which is shown in Figure 2(b).  The red dots correspond to the sites in our $q$-state Potts
model.  These red dots are then connected by green arcs, forming
a network as shown in Figure 2(c).  Figure 2(d) is the network extracted from Figure 2(c).  
We can further exploit the interesting properties of this example.  Figure 3 is the 
distribution of the number of neighbors that each of the current 449 precincts in Taipei 
city has.  We can call this the 
{\it connectivity} of a precinct.   We can also approximate this by a Gaussian Distribution
with a mean of about 4.6 and a standard deviation about 1.  This further supports the 
introduction of an interaction term among different sites in our discussion above.  The 
preferred voter district boundaries here will cut through those precincts with fewer 
connections to avoid long and thin precincts (units) within a voter district in order to satisfy 
compactness.  In fact, precincts with more connections will lie closer to the 
{\it center} of a voter district while precincts with few connections will stay near or at the
district boundaries.  

%Figure 4
\begin{figure}
\includegraphics[width=10cm]{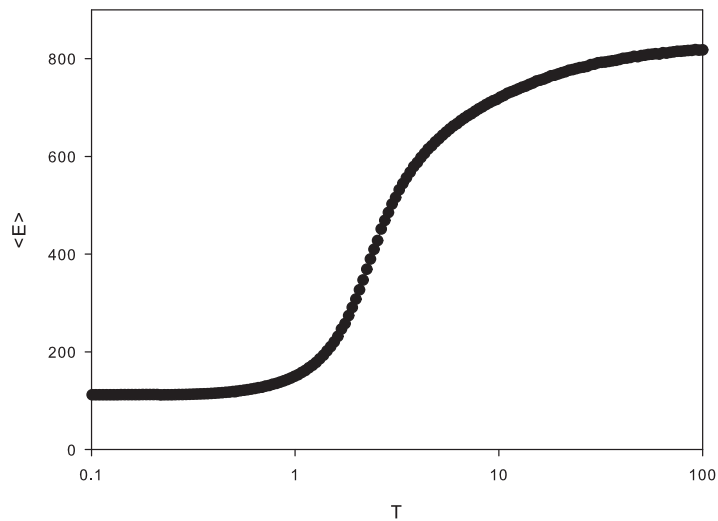}% Here is how to import EPS art
\caption{Energy of the system vs. temperature.}  
\end{figure}

%Figure 5
\begin{figure}
\includegraphics[width=10cm]{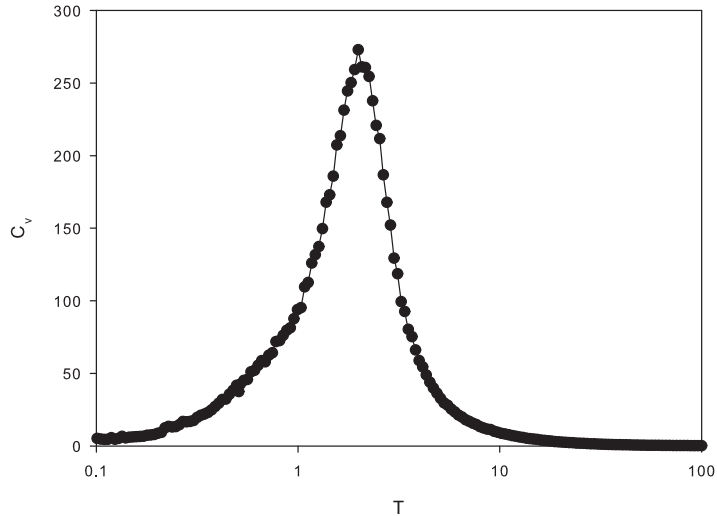}% Here is how to import EPS art
\caption{Specific heat capacity $C_V$ of our system vs. temperature.}  
\end{figure}

The competition between the two terms in the Hamiltonian in Eq. (2.11) defines the 
statistical properties of the model.  Figure 4 shows the energy of the system vs. 
temperature.  As we can see, the system undergoes a phase transition around 
$T = 2.0$ with an energy $E \approx 308$, when simulated $\lambda_P$ and
$\lambda_D$ are equal to 50 and 1 respectively.  The phase transition here 
corresponds to the formation of domains, or the aggregation of precincts into 
compact voter districts.  As mentioned above, the total number of sites in this 
system is 449 while $q$ is equal to 8, the number of precincts and voter districts 
in Taipei respectively.  To make sure there 
is no peculiar behavior in our system, we have actually started from $T = 10^7$ and 
gradually lowered the temperature and obtained a smooth curve all the way down to 
$T = 0.1$ where there is no further change in the ground state energy.  

Figure 5 is a plot of the specific heat capacity $C_V$ vs. temperature.  One can see that
there is a peak around temperature, $T = 2.0$, with a peak value of about 273, 
which is an indication that this is
a phase transition, where large domains are formed and condensed and the temperature
at which the peak appears corresponds to the critical temperature.  

Since our example here is a finite system (with only 449 sites), finite size effect prevents 
the peak to blow up.  In order to see that this model has a phase transition, we construct
an artificial system which has similar interaction terms as our example but with a 
varying number of sites to demonstrate the critical properties in the thermodynamic limit.  
The artificial model is our $q$-state Potts model on a periodic two dimensional triangular 
lattice with a Hamiltonian of the form similar to Eq. (2.11).  We again set $q = 8$ in this 
system and assume Gaussian distributions in both the voter population and the number of
connections for the sites.  Again, we normalize the 
total voter population to 1, independent of the number of sites.  For the connections, 
since each site on the triangular lattice could at the most connect to six nearest neighbors,
we normalize to have the peak of the Gaussian distribution to be at 3, and the cutofff 
at 6.  Figure 6 is a plot of the energy of the artificial 
system vs. temperature.  In the figure, we have included the energy of latttice sizes from 
$8 \times 8$ to $64 \times 64$.  One can clearly see the sharpening of the transition 
around $T \approx 1.6 $, with $E$ about 990, again with $\lambda_P$ and
$\lambda_D$ are equal to 50 and 1 respectively.  

%Figure 6
\begin{figure}
\includegraphics[width=10cm]{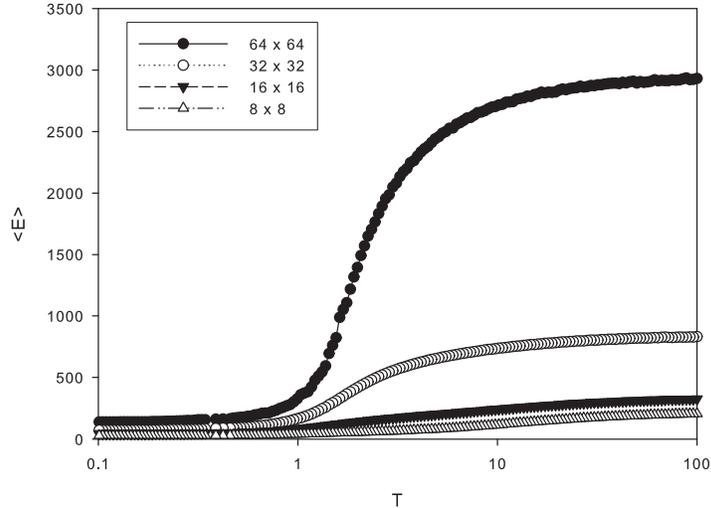}% Here is how to import EPS art
\caption{Energy of the artificial system vs. temperature.}  
\end{figure}

%Figure 7
\begin{figure}
\includegraphics[width=10cm]{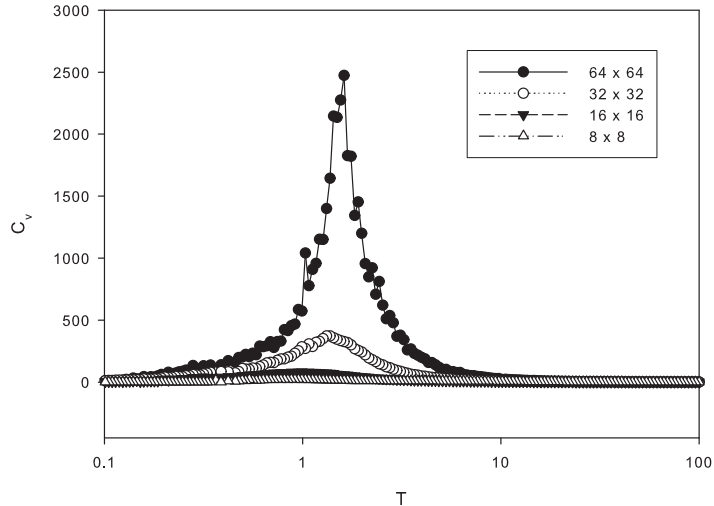}% Here is how to import EPS art
\caption{Specific heat capacity $C_V$ of the artificial system vs. temperature.}  
\end{figure}

Figure 7 is a plot of the specific heat capacity $C_V$ vs. temperature of the system.  
We can here see clearly that $C_V$ will diverge as the size of the lattice grows, which 
confirms a phase transition of the system.  The peak value of $C_V$ for the lattice
$64 \times 64$ is about 2470.
 
Figure 8(a) is a map of the Taipei city and its 449 precincts. 
As one lowers the temperature, the system will eventually reach its ground state.  
In reality, one cannot have absolute voter population equality for each of the voter
districts.  The number of near optimal solutions will increase if one allows the 
percentage difference of voter population of a district from the average voter
population/district to increase.  Figure 8(b) is an illustration of the voter districting 
with the lowest ground state energy from our simulation with $\lambda_P$ and 
$\lambda_D$ equal to 50 and 1 respectively.  There are a total of 8 voter districts, 
drawn in different colors.

%Figure 8
\begin{figure}
\includegraphics[width=12cm]{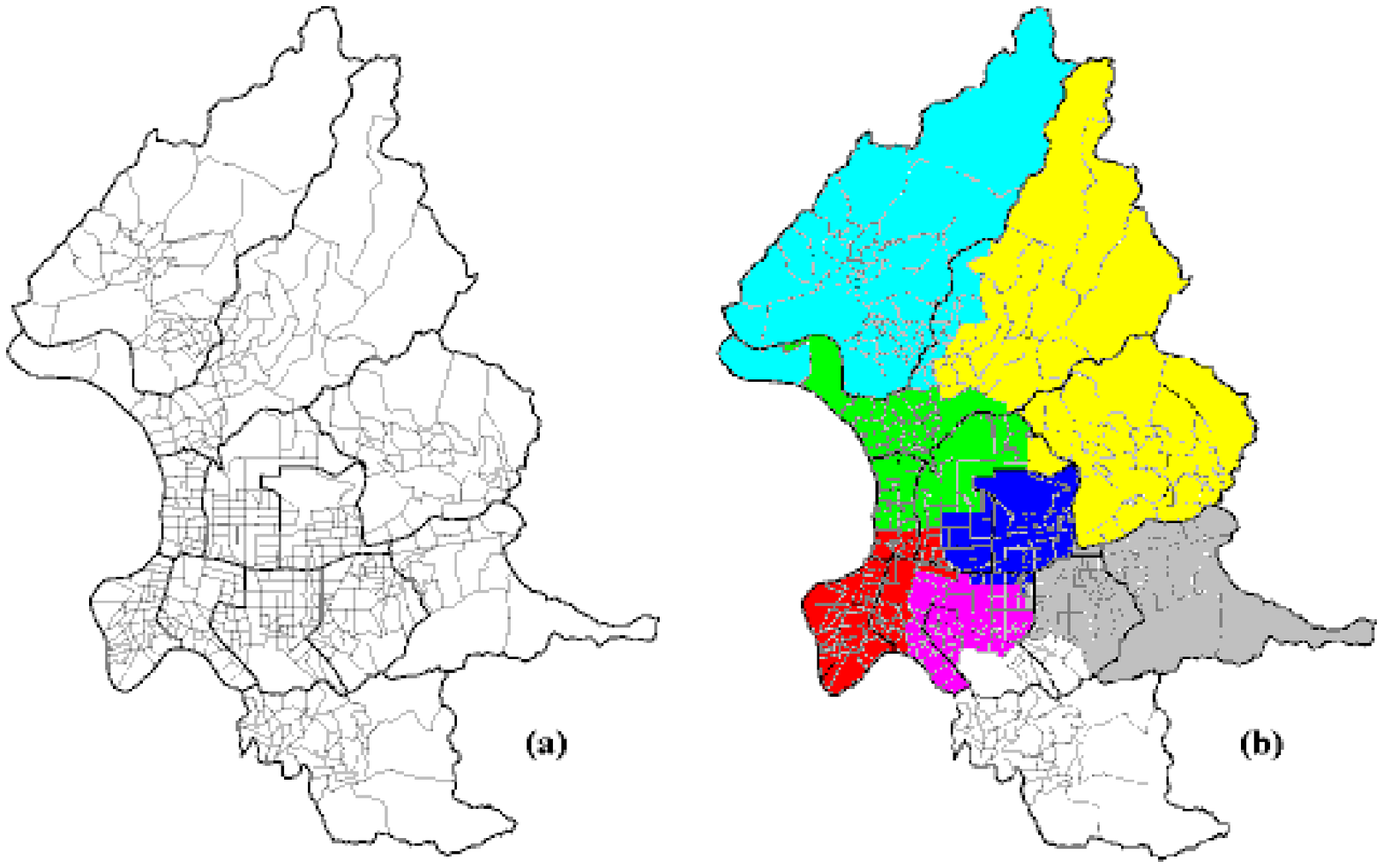}% Here is how to import EPS art
\caption{ (a) A map of the Taipei city and its 449 precincts; 
(b) an illustration of the 
voter districting in lowest energy state from our simulation 
with $\lambda_P$ and $\lambda_D$ equal to 50 and 1 respectively.  The 8 voter 
districts are drawn in different colors.  }  
\end{figure}

\hskip 1 cm {Table 1.  Voter districting in Taipei city with different $\lambda_A$
and $\lambda_D$.  }
\begin{center}
\begin{tabular}{|c|c|c|c|c|c|c|} \hline
$\lambda_P$  &  $\lambda_D$  & $E_{min}$  &  $E_P$  &  $E_D$  &  
$\Delta P_{max}$  &  $\Delta P_{max}/ \langle P \rangle$ \\ \hline
1  &  1  &  86.715  &  4.715  &  82  &  390925  &  1.5693  \\ \hline
5  &  1  &  97.073  &  1.615  &  89  &  143919  &  0.5778  \\ \hline
10  &  1  &  104.075  &  0.507  &  99  &  47145  &  0.1893  \\ \hline
50  &  1  &  110.537  &  0.031  &  109  &  1765  &  0.0071  \\ \hline
100  &  1  &  111.998  &  0.020  &  110  &  1846  &  0.0074  \\ \hline
500  &  1  &  145.778  &  0.014  &  139  &  801  &  0.0032  \\ \hline
\end{tabular}
\end{center}

Table 1 is the optimal energy state ($E_{min}$) that we obtain with
different values of $\lambda_P$ and $\lambda_D$.  $E_P$ and $E_D$ are the
contributions to $E_{min}$ from the voter population ($\lambda_P H_P$) and
district boundary ($\lambda_D H_D$) in Eq. (2.11).
The voter population of each 
precinct is taken from the 2004 Legislature Election.  Also included is the largest
deviation ($\Delta P_{max}$) of the voter population of a district from the average 
voter population $\langle P \rangle$ and the ratio 
($\Delta P_{max}/ \langle P \rangle$).
In the 2008 Legislature Election, the Central Election Commisson (CEC) of Taiwan
constrains $\Delta P_{max}/ \langle P \rangle$ of Taipei city to be less than 15\%.  
On the other hand, one also wants to have a near minimal $E_D$ in order to 
guarantee contiguity and compactness.  Taking these into consideration,  
$\lambda_P$ somewhere between 10 and 100 is preferred for the districting.

\section{Summary and Discussion}

In this paper, we show how to use a statistical physics model to study a social economics
problem.  We have mapped the Political Districting Problem to a $q$-state Potts model 
in which the constraints can be written as interaction between sites or external fields acting 
on the system.  Districting into $q$ voter districts is equivalent to finding the ground 
state of the system.  Searching for an optimal solution for the ground state becomes an
optimization problem and standard optimization algorithms such as the Monte Carlo method
or simulated annealing method can be employed here.  

The system undergoes a phase transition as one lowers the temperature.  This 
transition can be understood as follows.  At high temperature, only small domains
are formed and the whole system is in a random state.  As the temperature decreases,
large domains will begin to form in order to lower the energy of the system.  At the
critical temperature, the system will form large domains and thus will
approach the ground state configuration.  

In the example above, we studied the 2008 Taiwan Legislature Election with two constraints,
viz. voter population equality and compactness.  With a suitable choice of the ratio between
$\lambda_D$ and $\lambda_P$, the near optimal solutions also satisfy the contiguity 
condition.  One can also add other interaction terms for extra constraints.  In our example
here, Taipei city itself has currently 12 administrative zones.  The CEC of Taiwan prefers to 
have no more than 2 administrative zones in each voter district.  
Hence the districting here corresponds to adding another constraint to the Hamiltonian.  
One can, for example, add another term to the Hamiltonian which takes the following form
$$
H_A =  \lambda_A \sum_{i,j,k} \delta_{S_i,S_j} \delta_{S_j,S_k} \delta_{S_k,S_i} 
(1 - \delta_{A_i,A_j}) (1 - \delta_{A_j,A_k}) (1 - \delta_{A_k,A_i}) \,\,\, ,
\eqno(4.1)
$$
\noindent
where $A_i$ here refers to the administrative zone that site $i$ belongs to and $i, j, k$ all
belong to the same voter district.  One can see that in (4.1), the right hand side will give a
finite contribution if the sites within a voter district belong to 3 or more administrative 
zones.  A large $\lambda_A$ practically eliminates such a possibility.

In general, one can include additional terms to the total Hamiltonian to take care of
new constraints and the methodology we give here should equally be applicable to other
districting problems.  We have thus shown how one can use statistical physics
approach to study \lq\lq socio-econophysics\rq\rq \, problems and demonstrate with the 
example of districting Taipei city in the 2008 Taiwan Legislature Election.

This work was supported in part by the National Science Council, Taiwan, R.O.C. 
(grant no. NSC-94-2112-M-001-019).


\begin{thebibliography}{99}
%\bibitem[\ast]{li}
%E-mail address: {\tt spli@phys.sinica.edu.tw} 
\bibitem{kai}
{H. Kaiser, Midwest Journal of Political Science, $\chi$ (1966). }
\bibitem{nag}
{S. Nagel, Stanford Law Review 17 (1965) 863. }
\bibitem{gar}
{R.S. Garfinkel and G.L. Nemhauser, Manage. Sci. 16 (1970) 495. }
\bibitem{geo}
{J.A. George, B.W. Lamar and C.A. Wallace, Proceedings of the Operational Research Society 
of New Zealand 29 (1993) 276. }
\bibitem{alt}
{M. Altman, Rutgers Comput. and Technical Law J 23 (1997) 81. } 
\bibitem{meh}
{A. Mehrotra, E.L. Johnson and G.L. Nemhauser, Manage. Sci. 44 (1988) 1100. } 
\bibitem{bou}
{J.P.Bouchard and M. Potters, {\it{Theory of Financial Risks}}(Cambridge University Press, 
Cambridge, 2001). }
\bibitem{man}
{R.M. Mantefna and H.E. Stanley, {\it{An Introduction to Econophysics}}(Cambridge University
Press, Cambridge, 2000). }
\bibitem{voi}
{J. Voit, {\it{The Statistical Mechanics of Financial Markets}} (Springer, Berlin, 2001). }
\bibitem{fu}
{Y. Fu and P.W. Anderson, J. Phys. A: Math. Gen. 19 (1986) 1605. }
\bibitem{pot}
{R.B. Potts, Proc. Camb. Phil. Soc. 48 (1952) 106. } 
\bibitem{you}
{H.P. Young, Legislative Studies Quarterly XIII (1988) 105. }
\end{thebibliography}
\end{document}